# Deep phenotyping in *C. elegans*

Nan Xu

*Abstract*

Deep phenotyping study has become an emerging field to understand the gene function and the structure of biological networks. For the living animal *C. elegans*, recent advances in genome-editing tools, microfluidic devices and phenotypic analyses allow for a deeper understanding of the genotype-to-phenotype pathway. In this article, we reviewed the evolution of deep phenotyping study in cell development, neuron activity, and the behaviors of intact animals.

I. **Introduction**

Phenotyping study aims to provide quantitative classifications of traits such as disease manifestations, and to discover the underlying causes from the interaction among genotype, environments, and stochastic effects [1-3]. Understanding these underlying causes significantly relies on morphological, functional or behavioral phenotypic changes upon genetic perturbations.

Since the nematode Caenorhabditis elegans (*C. elegans*) is collected in 1962, it has been used as a promising model organism to study animal development and behavior. As a small multi-cellular organism, it is transparent and the lifetime is short. Therefore, it is widely used in high-throughput imaging and various experiments. Traditional phenotyping study in *C. elegans*, however, has focused on traits that are identifiable by human vision. Therefore, the data collecting processes are usually time-consuming and labor intensive. A key challenge of the phenotyping studies in *C. elegans* is inevitably limited in the sense that phenotypes of individual worms cannot be matched to the genetic variations. Recent advances in biotechnology, microfluidic devices, and image processing provide promising tools for modern phenotyping study. In particular, new technology advances enable high-throughput automated imaging of *C. elegans*, and provide highly specified and comprehensive phenotypic results. Most importantly, new technologies have given birth to an emerging field called *deep phenotyping*, whose goal is to discover pathways from genetic variations to phenotypes. In this article, we review the recent development of the new technology advances in phenotyping study of worms (described in Section II), as well as the evolution of deep phenotyping study in cell development (Section III), neuron activity (Section IV), and behaviors of intact animals (Section V), respectively.

### I. Technologies that empowers the phenotyping study

Advances in biology techniques, hardware devices, and mathematical analysis, have enabled high-throughput imaging and analysis of complex phenotypical traits of *C. elegans*. Please see **Figure 1** for an illustration of a typical imaging workflow of phenotyping study.

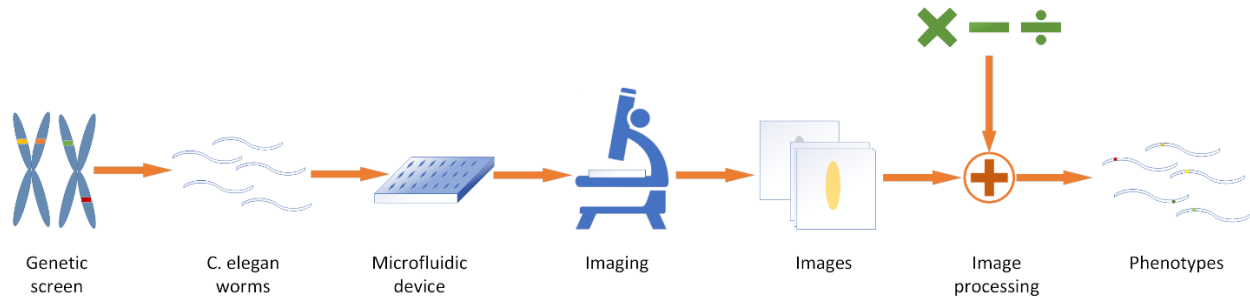

**Figure 1: Overview of imaging workflow for the phenotyping study.**

First of all, genetic screening (as illustrated in Figure 1), a key biotechnology, provides a 'phenotype-to-genotype' approach which links phenotypic traits to their genotypic causes including specific genes, RNAs, or alleles. The traditional forward genetic screening, which screens the mutagenized populations, has been widely used in phenotyping studies of cells, worms (e.g., [4, 5]), and other lab animals. In parallel, the reverse genetic screening, such as RNAi screening [6-9], has become prevalent for comprehensive genome-wide analysis in cell biology and development.

Secondly, microfluidic devices and hydrogel chips, which enable multi-worm loading and imaging, have been widely used as a high-throughput experimental platform for neuronal imaging [5, 10, 11] and behavioral analysis [10, 12, 13]. Among all microfluidic devices, the most popular ones include the multi-chamber device, the worm sorting device, and the worm arena (see **Figure 2**). The use of different types of microfluidic devices for high-throughput experiments of *C. elegans* has been reported in details in [11, 12, 14].

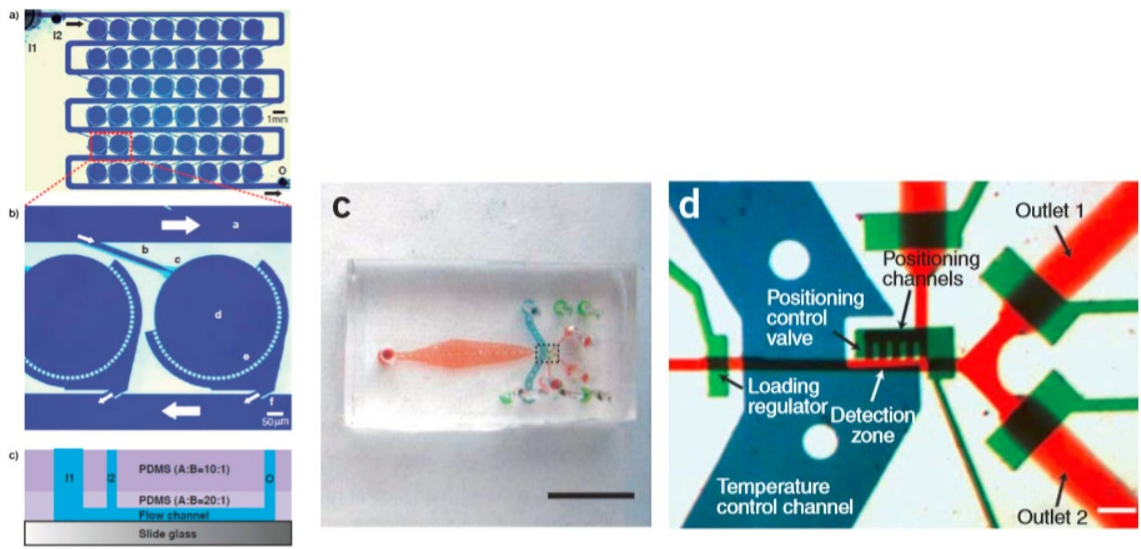

(a) Microfluidic chamber      (b) Sorting device

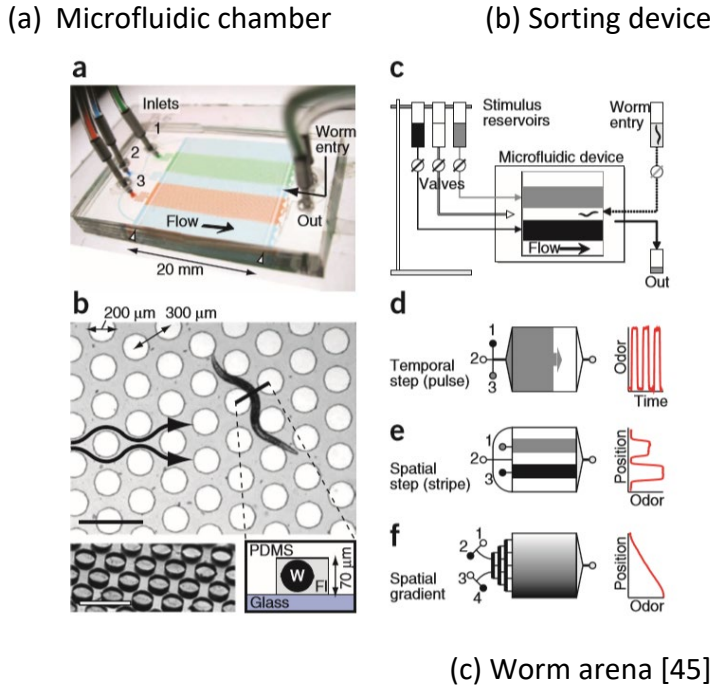

(c) Worm arena [45]

**Figure 2**. **Various microfluidic devices used for high-throughput imaging**

The third and the most important advance, which empowers the recent deep phenotyping studies, comes from the integration of mathematical modeling and quantitative analysis into image processing [15-19]. Mathematical models have been used to model various types of the biological phenomenon in 1) cell development, 2) neuron activity and 3) behavioral of intact animals. Specific models used in each of the three applications are discussed in sections III, IV and V, respectively. In parallel, statistical analysis has been widely used to quantify the numerical results for all above three applications. For instance, hierarchical clustering and K-mean clustering algorithms are used to extract the synaptic or behavioral features in order to identify mutants of *C. elegans* [5, 15-17]. A clustering algorithm is also introduced to identify regular T cells from cell populations [18]. In addition, statistical tests are used to quantify similarities between worm populations (e.g., [5, 20]).

II.       **Cell development**

Phenotyping studies of cellular development of C elegans include 1) the study of neuronal or synaptic development regulated by genes [4, 5, 9, 21-25], proteins [4, 26-28], and ageing and sex [4], and 2) the study of multicellular organisms such as living embryos [29, 30]. The evolution of phenotyping study in cellular development is described in the remainder of this section.

Tranditional approaches profile or quantify the visible phenotypes from microscopic images in low throughput. Among all these works, early study [4, 21, 22, 26], using forward genetic technology, focuses on identifying limited number of underlying causes, i.e., specific genes or proteins, in response to neuronal development. For example, the subset of neuron-specific genes is identified by profling the expression of sensory neuronal types [22]; several genes which regulate insulin secretion and lifespan are identified by profiling synaptic proteins [21]; specific genes that cause the neurodegeneration [4] and GABAergic neuronal development [26] are also identified. Later works [9, 23], with the aid of genome-wide forward and reverse genetic screening, are able to expand such investigation from the limited number of genes to the entire genome. In particular, 245 effective genes and 161 mutants are identified to play a role on cell fate from an embryo to an ASEL neuron in [23] and [9], respectively. Despite this progress, the number of worms used in imaging and analysis in above studies is still limited --- usually less than 20 in each experimental setting, and only visible phenotypic traits such as the presence of neuodegeneration, the expression of neuron types are identified, or some visible features including the intensity of synaptic puncta, the number of vesicles, the number of neurite swelling, and the length, area, and density of neurons are quantified.

To overcome the above two limitations, microfluidic devices are introduced to this field to improve the throughput of imaging; furthermore, image processing techniques are adoped to precisely identify the subtle and invisible phenotypic variations. The pioneer studies [20, 25] successfully integrate these two technologies. Specifically, by using a microfluidic sorting device in [25]  (Fig. 2(b)), 1) the worm on the chip is being imaged, and 2) the image is being quantified based on cellular phenotypes, and 3) the worm is sorted into the desired channel based on the quantification result. In this system, hundreds of worms can be continously imaged, processed, and sorted in real time. Inspired by this on-chip imaging system, an online image processing algorithm based on the supervised learning technique is developed in [20] to classify the subtle morphological features of synapses in real-time.

Inspired by the above technologies, the most recent studies (e.g., [5, 24]) further expand the exsiting analysis on big data to classify much more complex phenotypic traits. The first deep phenotyping study on synaptic formation, which adopts the automated sorting system in [20, 25], is introduced in [5]. In this study, a stepwise logistic regression model is used to separate the mutants from the wild type, and hiearchical clustering is used to identify the similarity of synaptic patterns between different mutants and between the mutant and the wild type (**as shown in Figure 3**). As a result, multidimentional traits of alleles and subtle mutations, which have been overlooked by traditional methods, are identified. These mutations were further validated by the intact animal observations as behavior defects. The other deep phenotyping study of *C. elegans* focuses on the transgenic effects on the morphology of neurons [24]. To enable the high-throuput imaging, a hydrogel based multiwell device is used to immobilize the worms. A computer-assisted microinjection system is adapted for the transgenesis. Phenotypeic traits including dendrite length, morphology and position of 78 mutant neuron classes are analyzed, and a vector library of neuron cell-specific markers of plasmids is generated by this study.

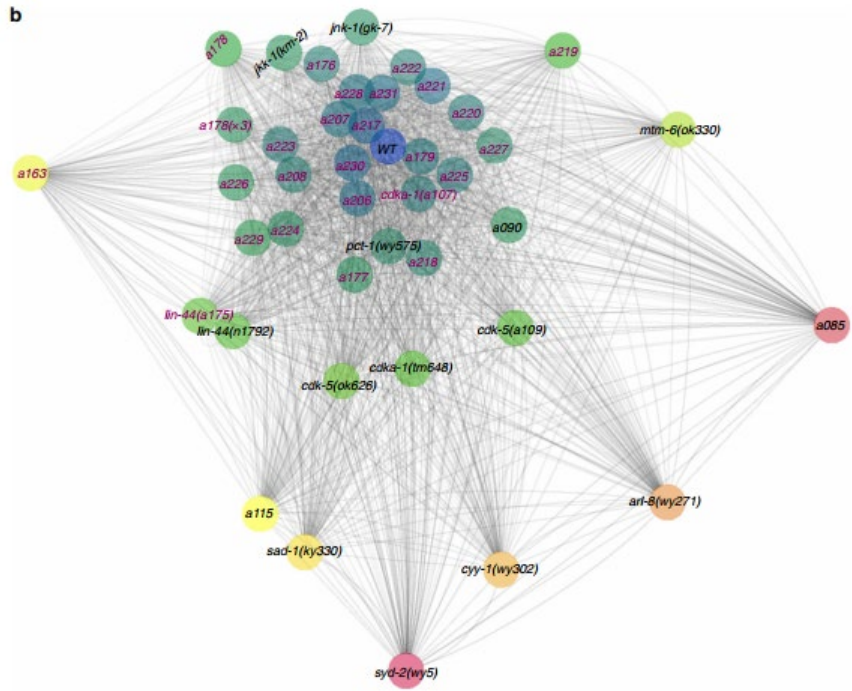

**Figure 3**: **Deep phenotyping reveals mutant relationships and differences in the phenotypic spectrum**

Comparing to the above studies on single cell formation, phenotyping study on living embryos needs to consider both the spatial and temporal factors, and therefore it is challenging. A microfluidic array device developed in [29] provides an on-chip worm culture platform, to incubate and to synchronize embryo populations. Using this system, the morphogenesis and mitochondrial biogenesis of living embryos are studied. The variation among the embryo population during the full embryonic development provides the quantification of temporal phenotypes.

III. **Neuron activity**

The second prosperous area of phenotyping studies has been in the functional activity of neurons in response to various causes, including the locomotion or/and chemical stimulation [10, 31-34], environmental variability [35], learning [36], optogenetic manipulations [34, 37-41]. In most of these studies, the flourecent intensity of neurons is considered as the primary phenotypic trait whereas the corresponding behavioral patterns (e.g., forward, backward and pirouette movements), are considered as the secondary phenotypic trait. While the intact animals are considered in the experiment, either one single worm [10, 32, 34, 38] or a group of worms is imaged at a time [31]. The development of the phenotyping study in this field is detailed in the remainder of this section.

For the studies with single worm experiments, most of them have been focusing on identifying a limited number of neurons as the dedicated encoders of specific sensory inputs or motor outputs and studying the isolated sensory-to-motor pathways [10, 33, 34]. By using the whole brain Calcium imaging, recent studies [32, 33] greatly expand the number of neurons to be investigated from less than 5 to more than 100. In particular, Kato and his collegues [32] explore the distributed or population dynamics in *C. elegans*. The dynamics of brain state are recorded at different motor state and crawling speed of freely moving worms, as well as at various sensory stimulations. The later work [33] improves the imaging system and the behavioral analysis algorithm in [32] to track the whole brain neuron activities of freely moving animals (e.g., *C. elegans* and *Drosophila* larva) in a much higher spatial and temporal resolution.

Other studies with single worm experiments use noninvasive approach to interfere the intracellular signaling, nervous system and the behavior of lab animals [41-43]. In particular, optogenetics are used to excite and silent the cells of interests. For example, a multimodal illumination system is introduced in [34], which optically manipulates sensory and command neurons in the *C. elegans* to study their roles in behaviors of freely moving animals. Other works (e.g., [38, 40]) focus on the role of pre- versus post- synaptic defects and synaptic transmission in light evoked behavior phenotypes and postsynaptic currents. More details of the studies which use the optogenetics technique are reported in the review paper [40].

The multi-worm expeirments, which inspect the neural activities and behaviors of multple worms simultanously, are firstly enabled by the microfluidic arena in [31]. Specifically, the neuronal responses to multiple odors, odor concentrations, and temporal and spatial patterns, and responses to pharmacological manipulation for both anethetized and up to 20 freely moving worms are tracked. Correlation between chemisensory neruon (e.g., AWA, AWC) and interneuron (e.g., AYA, AIA, AVA) activities as well as the behavior patterns (e.g., forward, backward and pirouette movements) indicate functional excitatory and inhibitory interactions in the local neuron network.

## IV.     Behavior of intact animals

The third area of phenotyping study is on behaviors of intact animals. This area has evolved into a large body of literature in the last two decades. Comparing to the phenotyping studies on cell development and on neuron activity, the phenotyping study on behaviors of intact animals is more complex, as it aims to analyze the phenotypic features in both spatial and temporal dimentions. In the remainder of this section, we describe 1) different types of behavioral patterns that have been studied in the literatures, 2) current mathematical models used to characterize behavior patterns, 3) existing software for tracking the behaviors, and 4) deep behavioral phenotyping studies by using the above technologies.

1. Various behavior phenotypes of C. elegan worms:

Various behavioral patterns, triggered by environmental perturbations, have been a subject of intense study. Phenotypes of chemotaxis [35, 44-51] have been the biggest focus in this area. The second biggest focus in this area includes phenotypes of thermotaxis [17, 52], the tap habituation [45], and gas/humidity vibrations [53, 54]. Other focuses include behaviorial phenoytpes motivated by aging or proteotoxicity associated phesiological decline [50, 55, 56]. More recent works aim to quantify and classify all possible behavioral features of the locomotion and the postures of C. elegan worms [19, 57-60].

Traditional experiments aim to observe the chemotaxis of the crawling worms. The early study [46], which reveals that pirouettes rather than the speed and the turning rate play the central role in chemotaxis, records the behavior patterns (i.e., the instantaneous position, speed, and turning rate) during chemotaxis over time on an agar plate. However, the spatial limitation of an agar plate restricts a high-content analysis of behaviors. To enable a high-throughput imaging and high-content analysis, [44] is the first study which uses a closed microfluidic device, the microfluidic arena (Figure 1 (c)), to study worm olfactory response. The microfluidic arena is optimized by various well-controlled stimulations. Analysis shows that olfactory behaviors can be dissected into known and new locomotory components, and identified genetic requirements for specific components of the olfactory response. The later work [35] further investigates the effects of learning on chemotaxis. More specifically, it investigates how the variability in food environment impact the future behavior in C. elegan worms and how a neural circuit decodes environmental variability to generate contextually appropriate decisions. A pair of sensor neurons ASI and ASK were found to encode variability in food concentration, which causes the downstream motor neurons to generate food searching behaviors. This study shows that learning modifies the same neuron driving behaviors. In addition to the traditional focus on the crawling worms, the study such as [47] focuses on the chemotaxis of swimming worms in microdroplet on a multi-chamber microfluidic device. Comparing to crawling worms on solid bacterial food, swimming worms exhibit more rapid responses to temporal odor variations. The multi-chamber microfluidic device includes multiple small circular chambers. Comparing to the microfluidic arena used in [31] and [44], the multi-chamber device provides a more closed and independent enviroment for each individual worm. Therefore, it allows for the analysis for more rapid and accurate behaviors for each individual.

A high-content multidimentional phenotyping method to study the thermal taxis of worms is introduced in [17]. A library of 47 mutated strains with various neuronal functional deficiencies is compared with the wild type strain (N2). Multiple aspects of the avoidance response of thermal stimuli at different intensities are quantified and profiled. Even though the combinations of molecules causing certain avoidance responses are unique, they may result in similar behavior phenotypes qualitatively. While the above studies are all based on the analysis of individual animals, [49] investigates the thermotaxis of a population of worms. Hundreds of worms are placed in a plate with linear thermal gradient, and are inspected simulatanously in the assay. The distribution and its statistics (i.e., the mean and the deviation) of the worm polutation are quantified. The population assay gives the advantage of extracting the behavior information in the statistical level. For more information on thermosensation phenotypes, please see the review paper [52].

2. Current mathematical models used to characterize behavior patterns

Various mathematical techniques have been used to model specific animal behavioral features and postures [35, 46, 60-66]. Three classes of mathematical models have been developed for different purposes which are discussed in the remainder of this subsection.

To model the behaviors and their transitions, stochastic processes have been popularly used (e.g., [57, 62, 63, 67]). Markov chain, in particular, has been consistently considered in the past several decades (e.g., [57, 67]). Based on the observation of pirouettes in worm chemotaxis experiments, a stochastic point model based on pirouette is developed in [46] to simulate the worm chemotaxis. Results of pirouette model of chemotaxis are validated by experimental data. Later study such as [62] models the fast and slow state of omega-turn of swimming worm as a stochastic process of a weighted sum of two Poisson processes. Ref. [63] models the transition between behavior states of drosophila as a stochastic process of conditional probability. By using the information theoretical approach, the behaviors are organized into a hierarchical structure which characterizes the underlying behavioral states.

To model the decision making of infortaxis, maximum mutual information (MI) has been commonly used to distinguish the situation of having the food or not (e.g., [35, 61]). This measure is usually jointly used with the stochastic or statistical models. For example, in [35] a Markov model is used to model the on food behavior, and a maximum MI-based filter is used to identify the time points of on-food which are relevant to modifying off-food search. In a similar example [61], the probability of detecting odorants is characterized as a time dependent Bayesian model, and the MI between observing and not observing odorant hits is updated over time. At the same time, the likelihood of food source being present is computed. In this model, two food searching phases, the local food search and the global food search, are considered. When the likelihood of food source being present within the searching range reaches its minimum, the global food searching phase starts. The model prediction has been supported by experimental measurements.

To model the general behavior features of moving worms, orthogonal representations recognized as "eigenworms", are widely used [19, 57-59]. The superposition of these representations covers most of the features of locomotion. Specifically, to quantitatively describe the motion of worms, the pioneer work [57] discovers four representations (or "eigenworms"), of which the linear combination covers 95% of the shape variance of worms. Such quantitative description is validated by real data. The extension study [60] maps the "eigenworm" analysis to the trajectories of crawling worms on an agar plate. In particular, a model is developed to connect "eigenworm" analysis with the speed and curvature of the worm's trajectory. This model reveals the subtle phenotypic differences in movement between two defective mutant strains. The "eigenworm" analysis has been built into several animal tracking software systems (e.g., [45, 58]) to analyze the worm behaviors.

In addition to the above mathematical models, the animal tracking systems (e.g. [10, 58, 65, 66]) described in Section V.3 also use various mathematical models to extract the behaviorial features and trajectories. The general pipeline includes 1) animal spine/curvature extraction, 2) tracking, 3) motion related feature extractions (i.e., "eigenworms" or velocity), and so on.

3. Existing software for tracking the behaviors

With the call of rapid quantification of behavior phenotypes, several tracking systems for lab animals are developed. Single C. elegan worm tracking system includes [10, 58, 65] (**as shown in Figure 4 (a)**). With the requirement of high-throughput imaging and behavioral analysis, several multiple worms tracking and analytical systems are available. For example, [45] extends their single-worm tracking system to a multi-worm tracker (MWT). MWT tracks and rapid quantifies the traditional behavioral paradigms, such as the random walk, the chemotaxis, and the habituation. MWT has been widely used in several studies (e.g., [44]). A more comprehensive overview of worm tracking is in [68]. Other tracking systems include the Maggot Tracker [69, 70] for Drosophila, and the FIMTrack [66] for Drosophila larvae or *C. elegans*, and the multi-animal tracker [55] which is demonstrated on C. elegan worms and can be extended to various animals (e.g., worms, flies, zebrafish, etc.). The major drawback of the above tracking systems for multiple animals is that they cannot correctly identify the animals during and after collisions. In particular, tracking is terminated during collision and new identifications get assigned to collided worms. With the request of a continuous behavioral analysis before and after animal collisions, several new tracking systems (e.g., [71, 72]) are developed to identify every individual throughout the video. However, the accuracy of these identification algorithms is still limited.

4. Deep phenotyping study in behaviors of intact animals

The advances in the software, algorithms and devices enable the deep phenotyping on animal behaviors. Recent studies perform gene variation in worm and quantify the corresponding behavioral phenotypes. For example, [54] studies the behavior responses to environmental changes in $O(2)$ and $CO(2)$. By doing gene variations, subtle behavioral phenotypes of worms are monitored and quantified in the arena. For each strain, about 20~30 worms are studied. Two quantitative trait loci (QTL) and the corresponding genes are identified. Comparing to [54], [73] conducts a more comprehensive screening on genes and more quantitative profiling on locomotive behaviors. Without varying environmental conditions in experiments, 227 neuronal signaling genes with viable homozygous mutants were screened, over four thousand worms are tracked, recorded and quantified individually. As a result, 87 genes and 370 genetic interactions associated with movement defects are revealed. Other studies (e.g., [19, 58]), using the "eigenworm" characterization, develop a more comprehensive library of behavioral phenotypes for *C. elegans*. In particular, [19] develops a dictionary of behavioral motifs for both the wild type and the mutants, which enables a fine classification among mutants as well as between the wild type and the mutants. In [58], a much longer behavior tracking, more behavioral features (e.g., motion, posture, and path as well as the frequencies and intervals between relevant behaviors) and more mutants are monitored and investigated, which establishes a more extensive database of behavioral phenotypes based on 305 strains (**Figure. 4 (c)**).

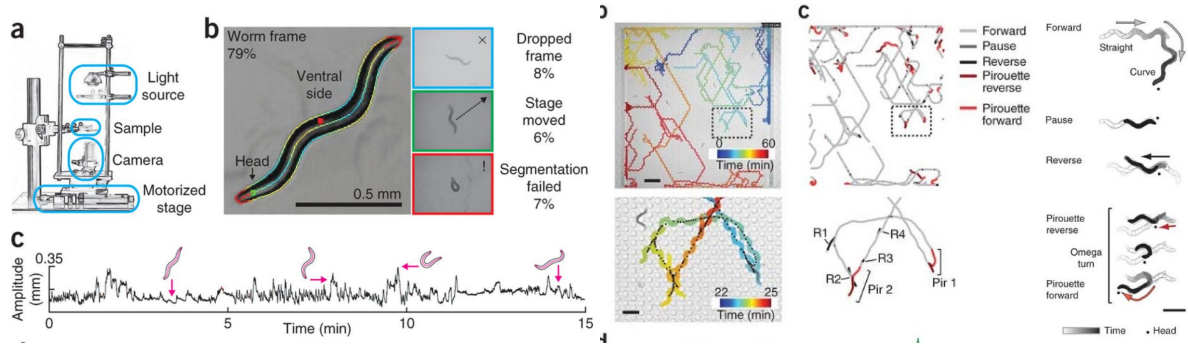

(a) Tracking individual worms [59]  (b) Tracking behaviorial phenotypes of multiple worms [45]

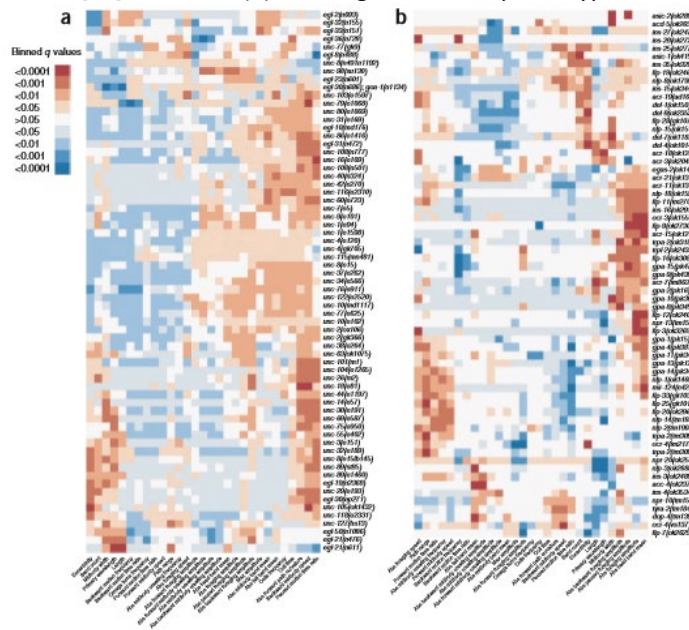

(b) Comparison of behavioral phenotypic features between N2 and mutants of *C. elegans*. Red values indicate features that have a significantly higher value in the mutant, whereas blue indicates significantly lower values in the mutant. Genes and features were both hierarchically clustered for easier comparison. [59]

**Figure 4. Behavioral phenotypes of *C. elegans***

## V. Discussion

In summary, the evolution of the phenotyping studies in the areas of cell development, neuron activity and the behavioral of intact animals is reviewed. The advanced technologies in genetic screening enable the investigation of a much deeper and broader underlying causes of various phenotypes in *C. elegans*.

While image-processing tools have been increasingly applied to perform quantitative analysis, this has been mostly useful for in vitro cell models. For live in vivo models (for example, genetic organisms), the number of markers (for example, fluorescent reporters) that can be simultaneously used is usually small, thus limiting the dimensions of the phenotype to be scored. There is also inherent complexity of working with intact animals, for instance, forward genetic screens in small model organisms are usually performed by phenotyping single animals, not populations of clones as in cell culture. These difficulties have prevented extensive use of large-scale highresolution image-based studies. In Caenorhabditis elegans, live phenotyping has mostly focused on drastic changes exhibited on gross features (for example, whole-animal or tissue-level changes). Most applications involving end point highresolution imaging are low dimensional, with the exception of live tracing of cell lineages and quantification of gene expression in embryos. However, the rich information encoded in fluorescence images of multicellular models has not been fully exploited at high resolution (that is, characterization of subcellular features within a living multicellular organism, thus missing the identification and characterization of phenotypic changes of weak alleles). Although different approaches have been used to identify chemically or genetically induced phenotypes, these have typically either screened for severe changes or have focused on behavioural or anatomical changes.

## V. Acknowledgement

The authors in this paper are grateful to Dr. Hang Lu, Mr. Tel Rouse for his suggestions on literatures in animal behaviors, and are thankful to Mr. Daniel Porto and Dr. Dhaval Patel for his helpful comments and discussions on the paper, and thankful to Ms. Kathleen Bates for her suggestions on literatures which include the figures of Microfluidic device.

## References

1. Furbank, R.T. and M. Tester, *Phenomics--technologies to relieve the phenotyping bottleneck.* Trends Plant Sci, 2011. **16**(12): p. 635-44.
2. Granier, C. and D. Vile, *Phenotyping and beyond: modelling the relationships between traits.* Curr Opin Plant Biol, 2014. **18**: p. 96-102.
3. Baynam, G., et al., *The facial evolution: looking backward and moving forward.* Hum Mutat, 2013. **34**(1): p. 14-22.
4. Hsu, J.M., et al., *Genetic analysis of a novel tubulin mutation that redirects synaptic vesicle targeting and causes neurite degeneration in C. elegans.* PLoS Genet, 2014. **10**(11): p. e1004715.
5. San-Miguel, A., et al., *Deep phenotyping unveils hidden traits and genetic relations in subtle mutants.* Nature Communications, 2016. **7**.
6. Bargmann, C.I., *High-throughput reverse genetics: RNAi screens in Caenorhabditis elegans.* Genome Biol, 2001. **2**(2): p. REVIEWS1005.
7. Echeverri, C.J. and N. Perrimon, *High-throughput RNAi screening in cultured cells: a user's guide*, in *Nat Rev Genet*. 2006: England. p. 373-84.
8. Boutros, M., F. Heigwer, and C. Laufer, *Microscopy-Based High-Content Screening.* Cell, 2015. **163**(6): p. 1314-25.
9. Poole, R.J., et al., *A Genome-Wide RNAi Screen for Factors Involved in Neuronal Specification in Caenorhabditis elegans.* PLoS Genet, 2011. **7**(6): p. e1002109.

10. Chronis, N., M. Zimmer, and C.I. Bargmann, *Microfluidics for in vivo imaging of neuronal and behavioral activity in Caenorhabditis elegans.* Nat Methods, 2007. **4**(9): p. 727-31.
11. Crane, M.M., et al., *Microfluidics-enabled phenotyping, imaging, and screening of multicellular organisms.* Lab on a Chip, 2010. **10**(12): p. 1509-1517.
12. Ben-Yakar, A., N. Chronis, and H. Lu, *Microfluidics for the analysis of behavior, nerve regeneration, and neural cell biology in C. elegans.* Current Opinion in Neurobiology, 2009. **19**(5): p. 561-567.
13. Chung, K., et al., *Microfluidic chamber arrays for whole-organism behavior-based chemical screening.* Lab on a Chip, 2011. **11**(21): p. 3689-3697.
14. Cornaglia, M., T. Lehnert, and M.A.M. Gijs, *Microfluidic systems for high-throughput and high-content screening using the nematode Caenorhabditis elegans.* Lab on a Chip, 2017. **17**(22): p. 3736-3759.
15. Zhang, S., et al., *Profiling a Caenorhabditis elegans behavioral parametric dataset with a supervised K-means clustering algorithm identifies genetic networks regulating locomotion.* J Neurosci Methods, 2011. **197**(2): p. 315-23.
16. Geng, W., et al., *Quantitative classification and natural clustering of Caenorhabditis elegans behavioral phenotypes.* Genetics, 2003. **165**(3): p. 1117-26.
17. Ghosh, R., et al., *Multiparameter behavioral profiling reveals distinct thermal response regimes in Caenorhabditis elegans.* BMC Biol, 2012. **10**: p. 85.
18. Kordasti, S., et al., *Deep phenotyping of Tregs identifies an immune signature for idiopathic aplastic anemia and predicts response to treatment.* Blood, 2016. **128**(9): p. 1193-205.
19. Brown, A.E.X., et al., *A dictionary of behavioral motifs reveals clusters of genes affecting Caenorhabditis elegans locomotion.* Proceedings of the National Academy of Sciences of the United States of America, 2013. **110**(2): p. 791-796.
20. Crane, M.M., et al., *Autonomous screening of C. elegans identifies genes implicated in synaptogenesis.* Nat Methods, 2012. **9**(10): p. 977-80.
21. Ch'ng, Q., D. Sieburth, and J.M. Kaplan, *Profiling synaptic proteins identifies regulators of insulin secretion and lifespan.* PLoS Genet, 2008. **4**(11): p. e1000283.
22. Colosimo, M.E., et al., *Identification of thermosensory and olfactory neuron-specific genes via expression profiling of single neuron types.* Current Biology, 2004. **14**(24): p. 2245-2251.
23. Sarin, S., et al., *Genetic screens for Caenorhabditis elegans mutants defective in left/right asymmetric neuronal fate specification.* Genetics, 2007. **176**(4): p. 2109-30.
24. Gilleland, C.L., et al., *Computer-Assisted Transgenesis of Caenorhabditis elegans for Deep Phenotyping.* Genetics, 2015. **201**(1): p. 39-46.
25. Chung, K., M.M. Crane, and H. Lu, *Automated on-chip rapid microscopy, phenotyping and sorting of C. elegans.* Nature Methods, 2008. **5**(7): p. 637-643.
26. Najarro, E.H., et al., *Caenorhabditis elegans flamingo cadherin fmi-1 regulates GABAergic neuronal development.* J Neurosci, 2012. **32**(12): p. 4196-211.
27. Withee, J., et al., *Caenorhabditis elegans WASP and Ena/VASP proteins play compensatory roles in morphogenesis and neuronal cell migration.* Genetics, 2004. **167**(3): p. 1165-76.
28. Sharifnia, P. and Y. Jin, *Regulatory roles of RNA binding proteins in the nervous system of C. elegans.* Front Mol Neurosci, 2014. **7**: p. 100.
29. Cornaglia, M., et al., *An automated microfluidic platform for C-elegans embryo arraying, phenotyping, and long-term live imaging.* Scientific Reports, 2015. **5**.
30. Banerjee, P., et al., *Deep phenotyping to predict live birth outcomes in in vitro fertilization.* Proc Natl Acad Sci U S A, 2010. **107**(31): p. 13570-5.
31. Larsch, J., et al., *High-throughput imaging of neuronal activity in Caenorhabditis elegans.* Proc Natl Acad Sci U S A, 2013. **110**(45): p. E4266-73.

32. Kato, S., et al., *Global brain dynamics embed the motor command sequence of Caenorhabditis elegans.* Cell, 2015. **163**(3): p. 656-69.
33. Venkatachalam, V., et al., *Pan-neuronal imaging in roaming Caenorhabditis elegans.* Proc Natl Acad Sci U S A, 2016. **113**(8): p. E1082-8.
34. Stirman, J.N., et al., *Real-time multimodal optical control of neurons and muscles in freely behaving Caenorhabditis elegans.* Nature Methods, 2011. **8**(2): p. 153-U78.
35. Calhoun, A.J., et al., *Neural Mechanisms for Evaluating Environmental Variability in Caenorhabditis elegans.* Neuron, 2015. **86**(2): p. 428-41.
36. Ha, H.I., et al., *Functional organization of a neural network for aversive olfactory learning in Caenorhabditis elegans.* Neuron, 2010. **68**(6): p. 1173-86.
37. Schultheis, C., et al., *Optogenetic Long-Term Manipulation of Behavior and Animal Development.* Plos One, 2011. **6**(4).
38. Liewald, J.F., et al., *Optogenetic analysis of synaptic function.* Nature Methods, 2008. **5**(10): p. 895-902.
39. Husson, S.J., A. Gottschalk, and A.M. Leifer, *Optogenetic manipulation of neural activity in C. elegans: From synapse to circuits and behaviour.* Biology of the Cell, 2013. **105**(6): p. 235-250.
40. Gottschalk, A., *Optogenetic Analysis of the Function of neural Networks and synaptic Transmission in Caenorhabditis elegans.* Neuroforum, 2014. **20**(4): p. 278-286.
41. Glock, C., J. Nagpal, and A. Gottschalk, *Microbial Rhodopsin Optogenetic Tools: Application for Analyses of Synaptic Transmission and of Neuronal Network Activity in Behavior*, in *C. Elegans: Methods and Applications, 2nd Edition*, D. Biron and G. Haspel, Editors. 2015, Humana Press Inc: Totowa. p. 87-103.
42. Zemelman, B.V., et al., *Selective photostimulation of genetically chARGed neurons.* Neuron, 2002. **33**(1): p. 15-22.
43. Boyden, E.S., et al., *Millisecond-timescale, genetically targeted optical control of neural activity.* Nat Neurosci, 2005. **8**(9): p. 1263-8.
44. Albrecht, D.R. and C.I. Bargmann, *High-content behavioral analysis of Caenorhabditis elegans in precise spatiotemporal chemical environments.* Nat Methods, 2011. **8**(7): p. 599-605.
45. Swierczek, N.A., et al., *High-throughput behavioral analysis in C. elegans.* Nat Methods, 2011. **8**(7): p. 592-8.
46. Pierce-Shimomura, J.T., T.M. Morse, and S.R. Lockery, *The fundamental role of pirouettes in Caenorhabditis elegans chemotaxis.* J Neurosci, 1999. **19**(21): p. 9557-69.
47. Luo, L., et al., *Olfactory behavior of swimming C-elegans analyzed by measuring motile responses to temporal variations of odorants.* Journal of Neurophysiology, 2008. **99**(5): p. 2617-2625.
48. Larsch, J., et al., *A Circuit for Gradient Climbing in C. elegans Chemotaxis.* Cell Rep, 2015. **12**(11): p. 1748-60.
49. Ito, H., H. Inada, and I. Mori, *Quantitative analysis of thermotaxis in the nematode Caenorhabditis elegans.* J Neurosci Methods, 2006. **154**(1-2): p. 45-52.
50. Dosanjh, L.E., et al., *Behavioral Phenotyping of a Transgenic Caenorhabditis elegans Expressing Neuronal Amyloid-beta.* Journal of Alzheimers Disease, 2010. **19**(2): p. 681-690.
51. Chai, C.M., C.J. Cronin, and P.W. Sternberg, *Automated Analysis of a Nematode Population-based Chemosensory Preference Assay.* Jove-Journal of Visualized Experiments, 2017(125).
52. Schafer, W.R., *Tackling thermosensation with multidimensional phenotyping.* BMC Biol, 2012. **10**: p. 91.
53. Vidal-Gadea, A., et al., *Caenorhabditis elegans selects distinct crawling and swimming gaits via dopamine and serotonin.* Proceedings of the National Academy of Sciences of the United States of America, 2011. **108**(42): p. 17504-17509.

54. McGrath, P.T., et al., *Quantitative mapping of a digenic behavioral trait implicates globin variation in C. elegans sensory behaviors.* Neuron, 2009. **61**(5): p. 692-9.
55. Itskovits, E., et al., *A multi-animal tracker for studying complex behaviors.* BMC Biol, 2017. **15**(1): p. 29.
56. Maulik, M., et al., *Behavioral Phenotyping and Pathological Indicators of Parkinson's Disease in C. elegans Models.* Frontiers in genetics, 2017. **8**: p. 77-77.
57. Stephens, G.J., et al., *Dimensionality and dynamics in the behavior of C. elegans.* PLoS Comput Biol, 2008. **4**(4): p. e1000028.
58. Yemini, E., et al., *A database of Caenorhabditis elegans behavioral phenotypes.* Nat Methods, 2013. **10**(9): p. 877-9.
59. Feng, Z.Y., et al., *An imaging system for standardized quantitative analysis of C-elegans behavior.* Bmc Bioinformatics, 2004. **5**.
60. Stephens, G.J., et al., *From modes to movement in the behavior of Caenorhabditis elegans.* PLoS One, 2010. **5**(11): p. e13914.
61. Calhoun, A.J., S.H. Chalasani, and T.O. Sharpee, *Maximally informative foraging by Caenorhabditis elegans.* Elife, 2014. **3**.
62. Srivastava, N., D.A. Clark, and A.D.T. Samuel, *Temporal Analysis of Stochastic Turning Behavior of Swimming C. elegans.* Journal of Neurophysiology, 2009. **102**(2): p. 1172-1179.
63. Berman, G.J., W. Bialek, and J.W. Shaevitz, *Predictability and hierarchy in Drosophila behavior.* Proc Natl Acad Sci U S A, 2016. **113**(42): p. 11943-11948.
64. Karbowski, J., et al., *Systems level circuit model of C-elegans undulatory locomotion: mathematical modeling and molecular genetics.* Journal of Computational Neuroscience, 2008. **24**(3): p. 253-276.
65. Cronin, C.J., et al., *An automated system for measuring parameters of nematode sinusoidal movement.* Bmc Genetics, 2005. **6**.
66. Risse, B., et al., *FIMTrack: An open source tracking and locomotion analysis software for small animals.* PLoS Comput Biol, 2017. **13**(5): p. e1005530.
67. Miller, G.A., *The cognitive revolution: a historical perspective.* Trends in Cognitive Sciences, 2003. **7**(3): p. 141-144.
68. Husson, S.J., et al., *Keeping track of worm trackers.* WormBook, 2013: p. 1-17.
69. Aleman-Meza, B., et al., *High-content behavioral profiling reveals neuronal genetic network modulating Drosophila larval locomotor program.* Bmc Genetics, 2017. **18**.
70. Aleman-Meza, B., S.K. Jung, and W. Zhong, *An automated system for quantitative analysis of Drosophila larval locomotion.* BMC Dev Biol, 2015. **15**: p. 11.
71. Winter, P.B., et al., *A network approach to discerning the identities of C. elegans in a free moving population.* Sci Rep, 2016. **6**: p. 34859.
72. Perez-Escudero, A., et al., *idTracker: tracking individuals in a group by automatic identification of unmarked animals.* Nat Methods, 2014. **11**(7): p. 743-8.
73. Yu, H., et al., *Systematic profiling of Caenorhabditis elegans locomotive behaviors reveals additional components in G-protein G alpha q signaling.* Proceedings of the National Academy of Sciences of the United States of America, 2013. **110**(29): p. 11940-11945.
74. Duan, Q., et al., *LINCS Canvas Browser: interactive web app to query, browse and interrogate LINCS L1000 gene expression signatures.* Nucleic Acids Res, 2014. **42**(Web Server issue): p. W449-60.
75. Jones, D.S., et al., *Profiling drugs for rheumatoid arthritis that inhibit synovial fibroblast activation.* Nat Chem Biol, 2017. **13**(1): p. 38-45.
76. Niepel, M., et al., *Profiles of Basal and stimulated receptor signaling networks predict drug response in breast cancer lines.* Sci Signal, 2013. **6**(294): p. ra84.

77. Schindelman, G., et al., *Worm Phenotype Ontology: Integrating phenotype data within and beyond the C. elegans community.* Bmc Bioinformatics, 2011. **12**.
78. Hu, Y., et al., *Gene2Function: An Integrated Online Resource for Gene Function Discovery.* G3 (Bethesda), 2017. **7**(8): p. 2855-2858.
79. Drysdale, R. and C. FlyBase, *FlyBase : a database for the Drosophila research community.* Methods Mol Biol, 2008. **420**: p. 45-59.
80. Mallon, A.M., A. Blake, and J.M. Hancock, *EuroPhenome and EMPReSS: online mouse phenotyping resource.* Nucleic Acids Res, 2008. **36**(Database issue): p. D715-8.
81. Morgan, H., et al., *EuroPhenome: a repository for high-throughput mouse phenotyping data.* Nucleic Acids Res, 2010. **38**(Database issue): p. D577-85.
82. Schofield, P.N., et al., *The mouse pathology ontology, MPATH; structure and applications.* J Biomed Semantics, 2013. **4**(1): p. 18.
83. de Abreu, D.A.F., et al., *An Insulin-to-Insulin Regulatory Network Orchestrates Phenotypic Specificity in Development and Physiology.* Plos Genetics, 2014. **10**(3).
84. Frei, A.P., et al., *Highly multiplexed simultaneous detection of RNAs and proteins in single cells.* Nat Methods, 2016. **13**(3): p. 269-75.
85. Vinayagam, A., et al., *Integrating protein-protein interaction networks with phenotypes reveals signs of interactions.* Nat Methods, 2014. **11**(1): p. 94-9.
86. Nielsen, U.B., et al., *Profiling receptor tyrosine kinase activation by using Ab microarrays.* Proc Natl Acad Sci U S A, 2003. **100**(16): p. 9330-5.
87. Maciag, K., et al., *Systems-level analyses identify extensive coupling among gene expression machines.* Mol Syst Biol, 2006. **2**: p. 2006.0003.
88. Ji, N., et al., *Feedback control of gene expression variability in the Caenorhabditis elegans Wnt pathway.* Cell, 2013. **155**(4): p. 869-80.
89. Collinet, C., et al., *Systems survey of endocytosis by multiparametric image analysis.* Nature, 2010. **464**(7286): p. 243-9.
90. Samuel, A.P.W. and N. Saha, *Distribution of Red-Cell G6pd and 6pgd Phenotypes in Saudi-Arabia.* Tropical and Geographical Medicine, 1986. **38**(3): p. 287-291.
91. Samuel, A.P.W., et al., *Quantitative Expression of G6pd Activity of Different Phenotypes of G6pd and Hemoglobin in a Sudanese Population.* Human Heredity, 1981. **31**(2): p. 110-115.
92. Saha, N., et al., *The Inter-Tribal and Intra-Tribal Distribution of Red-Cell G6pd Phenotypes in Sudan.* Human Heredity, 1983. **33**(1): p. 39-43.
93. Brown, C.S., P.C. Goodwin, and P.K. Sorger, *Image metrics in the statistical analysis of DNA microarray data.* Proc Natl Acad Sci U S A, 2001. **98**(16): p. 8944-9.
94. Yin, Z., et al., *A screen for morphological complexity identifies regulators of switch-like transitions between discrete cell shapes.* Nat Cell Biol, 2013. **15**(7): p. 860-71.
95. Lee, H.J., et al., *Proteomic and Metabolomic Characterization of a Mammalian Cellular Transition from Quiescence to Proliferation.* Cell Rep, 2017. **20**(3): p. 721-736.
96. Bakal, C., et al., *Quantitative morphological signatures define local signaling networks regulating cell morphology.* Science, 2007. **316**(5832): p. 1753-6.
97. Smith, A.M., et al., *Quantitative phenotyping via deep barcode sequencing.* Genome Res, 2009. **19**(10): p. 1836-42.
98. Dar, R.D., et al., *Screening for noise in gene expression identifies drug synergies.* Science, 2014. **344**(6190): p. 1392-6.
99. Wahlby, C., et al., *High- and low-throughput scoring of fat mass and body fat distribution in C-elegans.* Methods, 2014. **68**(3): p. 492-499.

100. Yin, Z., et al., *Using iterative cluster merging with improved gap statistics to perform online phenotype discovery in the context of high-throughput RNAi screens.* BMC Bioinformatics, 2008. **9**: p. 264.
101. Perlman, Z.E., et al., *Multidimensional drug profiling by automated microscopy.* Science, 2004. **306**(5699): p. 1194-8.
102. Steininger, R.J., 3rd, et al., *On comparing heterogeneity across biomarkers.* Cytometry A, 2015. **87**(6): p. 558-67.
103. Eyer, K., et al., *Single-cell deep phenotyping of IgG-secreting cells for high-resolution immune monitoring.* Nat Biotechnol, 2017. **35**(10): p. 977-982.
104. Veugelers, M., et al., *Comparative PRKAR1A genotype-phenotype analyses in humans with Carney complex and prkar1a haploinsufficient mice.* Proc Natl Acad Sci U S A, 2004. **101**(39): p. 14222-7.
105. Fassihi, H., et al., *Deep phenotyping of 89 xeroderma pigmentosum patients reveals unexpected heterogeneity dependent on the precise molecular defect.* Proc Natl Acad Sci U S A, 2016. **113**(9): p. E1236-45.
106. Wagner, E., et al., *Analysis of tubular membrane networks in cardiac myocytes from atria and ventricles.* J Vis Exp, 2014(92): p. e51823.
107. Green, R.A., et al., *A high-resolution C. elegans essential gene network based on phenotypic profiling of a complex tissue.* Cell, 2011. **145**(3): p. 470-82.
108. Hur, M., et al., *MicroCT-based phenomics in the zebrafish skeleton reveals virtues of deep phenotyping in a distributed organ system.* Elife, 2017. **6**.
109. Tracy, R.P., *'Deep phenotyping': characterizing populations in the era of genomics and systems biology.* Curr Opin Lipidol, 2008. **19**(2): p. 151-7.
110. Lanktree, M.B., et al., *Phenomics: expanding the role of clinical evaluation in genomic studies.* J Investig Med, 2010. **58**(5): p. 700-6.
111. Wong, H.R., et al., *Toward a clinically feasible gene expression-based subclassification strategy for septic shock: proof of concept.* Crit Care Med, 2010. **38**(10): p. 1955-61.
112. Kaplan, J.M. and H.R. Wong, *Biomarker discovery and development in pediatric critical care medicine.* Pediatr Crit Care Med, 2011. **12**(2): p. 165-73.
113. Westbury, S.K., et al., *Human phenotype ontology annotation and cluster analysis to unravel genetic defects in 707 cases with unexplained bleeding and platelet disorders.* Genome Med, 2015. **7**(1): p. 36.
114. Anthony, C., et al., *Cooperation between Strain-Specific and Broadly Neutralizing Responses Limited Viral Escape and Prolonged the Exposure of the Broadly Neutralizing Epitope.* J Virol, 2017. **91**(18).
115. Haring, R. and H. Wallaschofski, *Diving through the "-omics": the case for deep phenotyping and systems epidemiology.* OMICS, 2012. **16**(5): p. 231-4.